\documentstyle[12pt]{article}
\newcommand{\bq}{\begin{equation}}
\newcommand{\eq}{\end{equation}}
\newcommand{\ds}{\displaystyle}

\begin{document} \begin{flushright} PITHA 98/9 \\ cond-mat/9803269
\end{flushright}
  \begin{center} {\Large\bf Exact Partition Functions
for the \\[0.1cm] Primitive Droplet Nucleation Model
\\[0.11cm] in 2 and 3 Dimensions
 \\[2,5cm] }
{\large \bf H.A. Kastrup\footnote{E-Mail: kastrup@physik.rwth-aachen.de} \\
Institute for Theoretical Physics\\ RWTH Aachen \\[0.2cm] 52056 Aachen,
 Germany}
\end{center} \vspace*{1.0cm}
  {\large \bf Abstract} \\[0.2cm]
 The grand canonical partition functions
 for primitive droplet nucleation models with an excess energy
 $\epsilon_n = -\hat{\mu}\, n + \sigma \,n^{1-\eta}, \eta= 1/d$, for
 droplets of
 $n$ constituents
  in $d$ dimensions are calculated exactly in closed form in the cases $d=2$
 and $3$ for all (complex)
  $\hat{\mu}$
  by exploiting the fact that the partition functions obey
  simple linear PDE. \\ \\ \\

  \newpage
 The picture that 1st-order phase transitions are  initiated by the
 formation (homogeneous nucleation) of expanding droplets
  of the new phase within the old phase is a familiar and important
  one (see the reviews [1-13]). In its most primitive form the droplets are
 considered to
  consist of $n$ constituents (e.g.\ droplets of Ising spins on a lattice or
  liquid droplets of molecules etc.),
  the ``excess'' energy $\epsilon_n$ of which is given by a "bulk" term
 proportional to
  $n$ and a "surface" term proportional to $n^{1-\eta}, \eta =1/d$, where
  $d \ge 2$ is the dimension of the system:  \bq   \epsilon_n =
   -\hat{\mu} n +
  \sigma n^{1-\eta}~,~~\eta =1/d~.\eq In the case of negative Ising spin
  droplets formed in a background of positive spins by turning
  an external magnetic field
  $H$ slowly negative,  below the critical temperature, the coefficient
  $\hat{\mu}$ in Eq.\ (1) takes the form $\hat{\mu}=-2H$. For liquid
 droplets of $n$
  molecules condensing from a supersaturated vapour one has $\hat{\mu}= \mu
  -\mu_c$, where $\mu$ is the chemical potential and $\mu_c$ its critical
  value at condensation point. \\ Assuming the average number $\bar{\nu}(n)
 $ of
  droplets with
  $n$ elements to be given by a Boltzmann factor, \bq   \bar{\nu}(n)
 \propto e^{\ds -\beta
  \epsilon_n}~,~ \beta = \frac{1}{k_B T}~~, \eq and  that the droplets form
  a noninteracting dilute gas leads to the grand canonical
   potential $\psi_d $
  per spin or per
  volume \begin{eqnarray}
  \psi_d (\beta, t = \beta \hat{\mu}) & =& \ln Z_G  = p \beta =
   \sum_{n=0}^{\infty}e^{\ds  t n-x n^{1-\eta}}~,~ \\ & & t=\beta
   \hat{\mu},~ x=\beta \sigma~;~ p:\mbox{pressure}~, \nonumber
 \\ & & d\psi_d = -U d \beta +
    \bar{n}dt~~ . \end{eqnarray}
  (Physical reasons may require to start the sum (3) not at $n=0$ but at some
  finite $n_0 >0$. This does not affect the following conclusions and can
  easily be taken care of. It is mathematically convenient to start at $n=0$.)
  \\ The series (3) converges for $t\le 0$ only,  which follows e.g.\ from
the  Maclaurin-Cauchy integral criterium \cite{14}. In applications to
 metastable systems
  one is
  interested, however, in the behaviour of $\psi_d(t,x)$ for $t \ge 0$
 which calls for
  an analytic continuation in $t$ or in the fugacity $z=e^t$ \cite{15,16,17}.
   \\ The
   series (3) has recently been discussed  in connection with the
 canonical quantum  statistics of Schwarzschild black holes \cite{18,19,20}.
  \\ Notice that it obeys the linear PDE
   \bq \partial_t^{d-1}\psi_d = (-1)^d \,\partial_x^{d}
    \psi_d~~, \eq in particular  \begin{eqnarray}
    \partial_t \psi_2 &=& \partial_x^2 \psi_2\,~~~ \mbox{ for } d=2~,
    \\ \partial_t^2\psi_3&=& -\partial_x^3 \psi_3~ \mbox{ for } d=3~,
    \end{eqnarray}
  which will be helpful to find $\psi_d$ for $t\ge 0$, in the
  following especially for
  $d=3$. \\ Before discussing the exact continuation of $\psi_2$ and $\psi_3$
  into the complex
  $t$- or $z$-plane, let me recall the known saddle point approximations
  for $\psi_d$ if  $n$ is turned into a continuous variable and $t>0$
 \cite{16,5}:
   \bq \tilde{\psi}_d=
  \int_0^{\infty}dn \, e^{\ds \beta(\hat{\mu}\, n -\sigma \,
   n^{1-\eta})}~~. \eq
  Setting $n= (\sigma/\hat{\mu})^d u^d $ yields \bq \tilde{\psi}_d = d \left
  (\frac{\sigma}{\hat{\mu}}\right)^d \int_0^{\infty}du \,u^{d-1}e^{\ds \beta
  h(u)}~,~ h(u)=a(u^d-u^{d-1})~,~ a=\frac{\sigma^d}{\hat{\mu}^{d-1}}~. \eq
  The function $h(u)$ has extrema at $u=u_0 =0$ and at $u=u_1=1-\eta$ with
  $h(u_1)=-a\eta(1-\eta)^{d-1},~ h''(u_1)=a(d-1)(1-\eta)^{d-3}~.$ According to
  the standard analysis \cite{21} the integral (9) has for large $\beta$  the
  asymptotic expansion \bq \tilde{\psi}_d \sim
  (1-\eta)^{d/2}\sqrt{\frac{\pi\, d}{2\beta \, \hat{\mu}}} \left
  (\frac{\sigma}{\hat{\mu}} \right)^{d/2}
  e^{\ds -\beta a \eta(1-\eta)^{d-1}}(i+O(1/\beta))~. \eq Here
 the path in the complex
  $u$-plane goes from $u_0 =0$ to $ u_1$ and then parallel to the imaginary
  axis to $+i\infty$. Thus, only half of the associated Gaussian integral
  along the steepest descents contributes! The leading terms in (10) are
  purely imaginary! For $d=2$ and $3$ (here see also Ref.\ [16]) one has
   \begin{eqnarray} \tilde{\psi}_2^{(\infty)}(t,x)&=&i
  \frac{\sqrt{\pi}x}{2t^{\ds
   3/2}} e^{\ds -x^2/(4t)}~,\\ \tilde{\psi}_3^{(\infty)}(t,x)&=&
   i \frac{2\sqrt{\pi}x^{3/2}}{3t^2}e^{\ds-\frac{4}{27}\frac{x^3}{t^2}}~.
   \end{eqnarray}
   Whereas $\tilde{\psi}_2^{(\infty)}$ is a solution of the heat equation (6),
 $\tilde{\psi}_3^{(\infty)}$ is not an exact solution of the
   corresponding equation (7)!
    We shall see below how this is to be understood in terms
   of the exact $\psi_3(t,x)$.  \\  For $d=2$ the function $\psi_2(t,x)$
   has been determined exactly for $t>0$ in closed form in Ref.\ \cite{18} by
    following Lerch's observation \cite{22} that
   the relation \begin{eqnarray}
  e^{\ds -\sqrt{nx^2}}& =& \frac{|x|}{\sqrt{\pi}}\int_0^{\infty}dv
  e^{\ds -x^2v^2/4-n/v^2} \nonumber \\ &=&
  \frac{|x|}{2\sqrt{\pi}}\int_0^{\infty}\frac{\ds d\tau}{\ds \tau^{3/2}}
  e^{\ds -x^2/(4\tau)-n\tau}=: \int_0^{\infty}d\tau\hat{K}(\tau, x)
  e^{\ds -n\tau}  \end{eqnarray} converts the series(3) into a
  geometrical one under the integral sign. Here
 \bq \hat{K}(t,x) =
   \frac{x}{2\sqrt{\pi t^3}}e^{\ds -x^2/(4t)}=-2\partial_x K(t,x)~,\eq
 where \bq K(t,x)= \frac{1}{\sqrt{4\pi t}} e^{\ds -x^2/(4t)} \eq
 is the "heat kernel" which obeys the heat equation (6).
  So does $\hat{K}(t,x)$.\\
  Now the series (3) can be summed exactly for
  $t<0\;(x>0)$ and then continued:
  \begin{eqnarray} \psi_2(z=e^t, x)&=&
\int_0^{\infty}d\tau
  \hat{K}(\tau, x)\frac{1}{1-e^{\ds (t-\tau)}}   \\ & =&
  \frac{x}{2\sqrt{\pi}}\int_1^{\infty}du\;
  \frac{\ds e^{\ds -x^2/(4\ln u)}}{\ln^{3/2}u}\frac{1}{u-
  z}~~. \end{eqnarray}
    The last relation shows that $\psi_2(z,x)$ can be continued analytically
    into the whole complex $z$-plane except for a cut from $z=1$ to
    $z=\infty$.\\
    The discontinuity of $\psi_2$ across
   the  cut ($z$ real and $>1$) is given
  by
  \bq \lim_{\ds \epsilon\rightarrow
  0^+}[\psi_2 (z+i\epsilon)-\psi_2 (z-i\epsilon)]= 2\pi i \hat{K}(t,x)~,~~
   \eq and,
  if one approaches the cut  from above,
   the limit \[
 \lim_{\ds \epsilon\rightarrow
  0^+}\psi_2(z+i\epsilon,x),~ z \mbox{ real and }>1~,\] is
   no longer a real-valued function of $z$,
   but has a nonvanishing
  imaginary part
  \bq \Im [\psi_2(t,x)] = \pi
  \hat{K}(t,x)=\frac{\sqrt{\pi}x}{2t^{\ds 3/2}} e^{\ds -x^2/(4t)}~~, \eq
  which   agrees exactly with that of the saddle point approximation
   (11)! \\
  The real part $\Re [\psi_2 (t,x)]$ is given by the principal
   value integral \bq
  \Re [\psi_2(t,x)]= \mbox{p.v.} \int_0^{\infty}d\tau\, \hat{K}(\tau,x)
   \frac{1}{1-e^{t-\tau}}~. \eq
 In order to prepare for the method by which to determine
  $\psi_3$, let us arrive at the
 above result for $\psi_2$ in a different way: \\ As the heat equation (6)
is invariant under the scale transformation $t \to \lambda t, x \to \lambda^2
x, \lambda > 0,$ it has solutions of the type $g(y=x^2/t),$ where
 $g(y)$ obeys
the ODE \bq g''+(\frac{1}{4} + \frac{1}{2y})g'=0~. \eq
 The solutions  (up to additive and
  multiplicative constants) are (see Ref.\ \cite{23})
\bq g(y) = y^{-1/4}e^{\ds -y/8} w(k=-1/4, m=1/4;y/4)~,\eq where $w$ is any
solution of Whittaker's standard form of the confluent hypergeometric
equation \cite{24}. As we want to associate the solution $g$ with the
imaginary part $\Im[\psi_2]$ in (11), we take
$w(-1/4,1/4;y/4)=aW_{-\frac{1}{4},\frac{1}{4}}(y/4), a$: con\-stant, where
$W_{k,m}$ is Whittaker's ``W''-function which has the property \cite{24}
\bq u^{-1/2}e^{\ds -1/2 u^2} W_{-\frac{1}{4},\frac{1}{4}}(u^2)
=2\int_u^{\infty}dve^{\ds -v^2}\equiv 2 \mbox{Erfc}(u). \eq Here Erfc(u)
 is the complementary error function. Thus we get
 \bq g(y)= \mbox{const.\ Erfc}(\frac{1}{2} \sqrt{y})~. \eq As \bq \partial_t
 \mbox{Erfc}(\frac{1}{2}\sqrt{y})=\frac{x}{4t^{3/2}}e^{\ds -x^2/(4t)}
 =\frac{\sqrt{\pi}}{2}\hat{K}(t,x)~,\eq we see that we may arrive at the
 solution (16) by essentially folding the imaginary part $\partial_t g(y)$
 with the function $1/(1-e^{t-\tau})$. (Note: If $F(t,x)$ is a
 solution of the heat equation, then any derivative or integral of $F(t,x)$
 with respect
 to $t$ or $x$ is a solution, too!) \\ The normalization of $g(y)$ is
 fixed by observing that for $\lambda >0$ we have (by rescaling $\tau \to
 \lambda^2 \tau$) \bq \psi_2(t,\lambda x)=
 \int_0^{\infty}d\tau \hat{K}(\tau,x)\frac{1}{1-\exp (\ds t-\lambda^2
 \tau)}~, \eq
 and that therefore \bq \lim_{\lambda \to 0}\psi_2(t,\lambda x)
 =\frac{1}{1-e^{\ds t}}~,~
 \lim_{\lambda \to \infty}\psi_2(t,\lambda x)
 = 1~.\eq
 We are now ready to proceed in the same way for $\psi_3$: As the PDE
 (7) is invariant under the scaling \cite{19} $t \to \lambda t, x \to
 \lambda^{2/3} x$, the ansatz $g(y=x^3/t^2)$  leads to the
 ODE \bq
g'''+(\frac{4}{27}+\frac{2}{y})g''+\frac{2}{9}(\frac{1}{y}+\frac{1}{y^2})g'
=0~, \eq which has solutions (see  Ref.\ \cite{23}) \bq g'(y)=y^{-1}e^{\ds
-\frac{2}{27}y}
w(1/2,1/6;\frac{4}{27}y)~.\eq  The choice
$w=W_{\frac{1}{2},\frac{1}{6}}(\frac{4}{27}y)$ provides the desired result: \\
With \bq g(y)= \int_y^{\infty}\frac{d\eta}{\eta}e^{\ds -\frac{2}{27}\eta}
W_{\frac{1}{2},\frac{1}{6}}(\frac{4}{27}\eta)  \eq we have \bq \partial_t g(y)
= \frac{2}{t} e^{\ds -\frac{2}{27}y}
W_{\frac{1}{2},\frac{1}{6}}(\frac{4}{27}y)~, \eq which for large $y,$ i.e.
large $x$ or small $t$,
takes the asymptotic form \cite{24} \bq \partial_t g(y) \sim
\frac{4}{(\sqrt{3})^3}\frac{x^{2/3}}{t^2}e^{\ds -\frac{4}{27}x^3/t^2}~. \eq
Except for a factor $\sqrt{3\pi}/2$ this is the same as
$\tilde{\psi}_3^{(\infty)}$ in Eq.\ (12). \\ Let us, therefore, try the ansatz
\begin{eqnarray} \psi_3(t,x)&=&\int_0^{\infty}d\tau\hat{K}_3(\tau, x)
\frac{1}{1-e^{\ds
t-\tau}}~,\\ & & \hat{K}_3(\tau,x)= \alpha\,\frac{1}{\tau}
e^{\ds -\frac{2}{27}x^3/\tau^2}
W_{\frac{1}{2},\frac{1}{6}}(\frac{4}{27}x^3/\tau^2)~~ , \nonumber
 \end{eqnarray} where $\alpha$ is a normalization constant to be determined.
 \\ That the function
$\psi_3$ of Eq.\ (33)
is indeed the right one can be seen as follows:
 It is a solution of
Eq.\ (7), because by construction $\hat{K}_3(t,x)$ is  a solution of that
equation (the partial derivative of $\psi_3$ with respect to $t$ can be
 replaced
under the integral by the negative derivative with respect to $\tau$,
 followed by a
partial integration with repect to $\tau$ and observing that $\hat{K}_3$
vanishes at the boundaries, see Eqs. (32) and (39)). \\ Furthermore, the
imaginary part of $\psi_3$ coincides with that of Eq.\ (12).
Finally, the required boundary conditions can be fulfilled by an appropriate
 choice of $\alpha$: \\ Rescaling
$\tau \to \lambda^{3/2} \tau$ yields \bq \psi_3(t, \lambda x) =
\int_0^{\infty}d\tau\hat{K}_3(\tau, x)\frac{1}{1-\exp(
t-\lambda^{2/3}\tau)}~. \eq In order to have (see Eq.\ (3))
 \bq \lim_{\lambda \to 0}\psi_3(t,\lambda x)
 =\frac{1}{1-e^{\ds t}}~,~~~
 \lim_{\lambda \to \infty}\psi_3(t,\lambda x)
 = 1~,
 \eq  we need  \bq \int_0^{\infty} d\tau \hat{K}_3(\tau, x) =1~. \eq
 Because the relation \bq \int_0^{\infty}\frac{du}{u}e^{\ds -u}
W_{\frac{1}{2},\frac{1}{6}}(2u) =\frac{\sqrt{\pi}}{\cos(\pi/6)} =
 2\sqrt{\frac{\pi}{3}}~ \eq
 holds \cite{25},
the normalization (36) requires $\alpha=\sqrt{3/\pi}$, so that finally  \bq
\hat{K}_3(\tau,x)= \sqrt{\frac{3}{\pi}}\,\frac{1}{\tau}
e^{\ds -\frac{2}{27}x^3/\tau^2}
W_{\frac{1}{2},\frac{1}{6}}(\frac{4}{27}x^3/\tau^2)~~. \eq
Let me add that $\hat{K}_3(t,x)$ has no zeros on the positive  real
$\tau$-axis for  $x>0$ \cite{24} and therefore is strictly positive there.
Furthermore, for $u\to 0$ one has \cite{24} \bq W_{\frac{1}{2},\frac{1}{6}}(u)
\to \frac{\Gamma(1/3)}{\Gamma(1/6)}u^{1/3}(1+O(u)) +
\frac{\Gamma(-1/3)}{\Gamma(-1/6)}u^{2/3}(1+O(u))~,  \eq
which is of interest for the behaviour of $\hat{K}_3(t,x) $ in
 the limits $x \to0$ or t$ \to \infty $. \\ One can check the result (38) by
 expanding the factor $1/(1-e^{(t-\tau)})$ in the integrand of (33)
   for $t<0$ in a geometrical series and comparing the coefficient
   \bq f(x,n) = \sqrt{\frac{3}{\pi}}
  \int_0^{\infty}d\tau
  \frac{1}{\tau}
e^{ -\frac{2}{27}x^3/\tau^2}\;
W_{\frac{1}{2},\frac{1}{6}}(\frac{4}{27}x^3/\tau^2)\,e^{\ds -n\, \tau} \eq
of $\exp(n\, t)$ with the one - $\exp(-x\, n^{2/3})$ -
 of the original series (3): \\
  Because $f(\lambda x, n)=f(x, \lambda^{3/2} n)$,
 we have $f(x,n)=g(x\, n^{2/3})$ and it is sufficient to consider (40) for
 $x=1$. $f(n) \equiv f(x=1,n)$ can be interpreted as the Laplace transform
 of \bq h(\tau^2)= \frac{1}{\sqrt{\tau^2}}
e^{ -\frac{2}{27\,\tau^2}}\;
W_{\frac{1}{2},\frac{1}{6}}(\frac{4}{27\,\tau^2})~, ~~f(n)=
\int_0^{\infty}d\tau e^{\ds -n\, \tau} h(\tau^2)~. \eq In that case one has
 (see Ref.\
\cite{26}) \bq f(n)= \frac{1}{\sqrt{\pi}} \int_0^{\infty}
\frac{du}{u^2}e^{\ds -n^2\,u^2/4}g(1/u^2)~, \eq where $g(y)$ is the Laplace
transform of $h(t=\tau^2)$. \\ As \cite{27} \bq g(y) =\int_0^{\infty} dt
 e^{\ds
-y\,t} h(t) = \frac{4}{\sqrt{27}}\, K_{1/3}(4\sqrt{\frac{y}{27}})~~, \eq where
$K_{\nu}(z)$ is the modified Hankel function \cite{28}, we finally get
 from Eqs.\ (42) and (43) the desired result (compare Ref.\
\cite{29}):
\bq f(n)= \frac{1}{3 \pi}\int_0^{\infty}\frac{dv}{v^{3/2}} e^{\ds -n^2\, v}
K_{1/3}(\frac{2}{\sqrt{27\, v}}) = e^{\ds -n^{2/3}}~. \eq
 As (see Ref.\ \cite{30}) \bq
 W_{\frac{1}{2},\frac{1}{6}}(2z)=  \frac{\sqrt{2z}}{2}(W_{0,\frac{1}{3}}(2z)+
  W_{0,\frac{2}{3}}(2z)) = \frac{z}{\sqrt{\pi}}(K_{\frac{1}{3}}(z) +
   K_{\frac{2}{3}}(z))~, \eq
   $\hat{K}_3(\tau, x)$ may also be expressed by these
   special functions. \\
  Having determined the exact partition functions, one can now study the
 associated
thermodynamical properties, e.g. the behaviour of the
 magnetization which is essentially given
by the derivative of $\psi_d$ with respect to $t$, the details of the
metastabilities etc. However, improving the mathematics does not
eliminate the many shortcomings of the model concerning the physics it
 is supposed
to approximate (see the Refs.\ [1-13])! Nevertheless, it is
 always pleasing to
have exact solutions for nontrivial models.  \\ \\
 I thank G.\ Roepstorff and T.\ Strobl for
a critical reading of the manuscript and the former for drawing my
attention to Ref.\ [17].
 \end{document}